# The Possibility of Emersion of the Outer Layers in a Massive Star Simultaneously with Iron-Core Collapse: A Hydrodynamic Model

**V. S. Imshennik*, K. V. Manukovskii, D. K. Nadyozhin, and M. S. Popov**

*Institute for Theoretical and Experimental Physics, ul. Bol'shaya Cheremushkinskaya 25, Moscow, 117259 Russia*

**Abstract**—We analyze the behavior of the outer envelope in a massive star during and after the collapse of its iron core into a protoneutron star (PNS) in terms of the equations of one-dimensional spherically symmetric ideal hydrodynamics. The profiles obtained in the studies of the evolution of massive stars up to the final stages of their existence, immediately before a supernova explosion (Boyes *et al.* 1999), are used as the initial data for the distribution of thermodynamic quantities in the envelope. We use a complex equation of state for matter with allowances made for arbitrary electron degeneracy and relativity, the appearance of electron–positron pairs, the presence of radiation, and the possibility of iron nuclei dissociating into free nucleons and helium nuclei. We performed calculations with the help of a numerical scheme based on Godunov's method. These calculations allowed us to ascertain whether the emersion of the outer envelope in a massive star is possible through the following two mechanisms: first, the decrease in the gravitational mass of the central PNS through neutrino-signal emission and, second, the effect of hot nucleon bubbles, which are most likely formed in the PNS corona, on the envelope emersion. We show that the second mechanism is highly efficient in the range of acceptable masses of the nucleon bubbles ($\lesssim 0.01 M_\odot$) simulated in our hydrodynamic calculations in a rough, spherically symmetric approximation.

Key words: *plasma astrophysics, hydrodynamics, and shock waves.*

## INTRODUCTION

The root cause of supernova explosions in massive stars ($M_{\rm MS} \geq 10 M_\odot$) after the completion of their thermonuclear evolution is commonly assumed to be the gravitational collapse of their central iron cores with masses $M_{\rm Fe}$ in a narrow range ($1.2 M_\odot \leq M_{\rm Fe} \leq 2 M_\odot$); the lower limit is approximately equal to the Chandrasekhar mass of an iron white dwarf. To be more precise, iron-core collapse is followed by a number of hydrodynamic processes in the compact central part of the cavity left from the iron core.

Hydrodynamic spherically symmetric models for the gravitational collapse of iron stellar cores [see, e.g., the review article by Imshennik and Nadyozhin (1982)] showed a clear separation of the collapse into two stages. At the first stage, the inner iron core with mass $M_{i\rm Fe} \leq 1 M_\odot$ (Imshennik 1992) or, to be more precise, with mass $(0.6–0.8) M_\odot$ (Nadyozhin 1998) collapses homologically. The remaining outer iron core with mass $M_{e\rm Fe} = M_{\rm Fe} - M_{i\rm Fe} \cong (0.4–1.4) M_\odot$ in the above range of iron-core masses lags well behind the inner core in contraction parameters (density, etc.). In a rough approximation, it can even be assumed to maintain the hydrostatic equilibrium of the initial state. Since these issues were discussed previously (Imshennik and Zabrodina 1999; Imshennik and Popov 2001), the hydrodynamic behavior of the outer iron core and outer layers of a massive star may be called the problem of their post-shock accretion (Brown *et al.* 1992) onto the embryo of a protoneutron star (PNS) with a mass of about $1 M_\odot$ formed at the end of the first collapse stage. The second collapse stage of the matter surrounding this PNS embryo can be identified with this post-shock accretion. Literally, this implies that at the initial time, there is a collapsed central object with mass $1 M_\odot$ and its surrounding layers in an initially hydrostatic equilibrium. These initial conditions for the model of post-shock accretion considered below seem to be in qualitative agreement with hydrodynamic models for the collapse of iron stellar cores. They allow the hydrodynamic processes in the remaining onionlike part of the star surrounding the iron core in a massive star of arbitrary total mass ($M_{\rm MS} \geq 10 M_\odot$) to be included in the analysis. Here, we show that these processes do not result in an explosion of the outer stellar layers surrounding the iron core as during a supernova explosion ($\sim 10^{51}$ erg). However, in the presence of low-mass ($\leq 0.01 M_\odot$) nucleon bubbles, they cause the emersion of the outer layers with a relatively low energy and with a total mass of several $M_\odot$ up to the inner boundary of the helium shell.

*E-mail: imshennik@vxitep.itep.ru

## FORMULATION OF THE PROBLEM

### *The System of Equations*

In most cases, the hydrodynamic behavior of the envelope in a massive star can be described by the system of equations of ideal hydrodynamics. In our spherically symmetric case $\left(\frac{\partial}{\partial\theta}, \frac{\partial}{\partial\varphi}, g_\theta = 0, g_\varphi = 0\right)$, this system in spherical coordinates $(r, \theta, \varphi)$ is known to be

$$\frac{\partial \rho}{\partial t} + \frac{1}{r^2}\frac{\partial}{\partial r}(r^2 \rho V_r) = 0, \quad (1)$$

$$\frac{\partial \rho V_r}{\partial t} + \frac{1}{r^2}\frac{\partial}{\partial r}(r^2 \rho V_r^2) + \frac{\partial P}{\partial r} = \rho g_r, \quad (2)$$

$$\frac{\partial \rho E}{\partial t} + \frac{1}{r^2}\frac{\partial}{\partial r}(r^2 V_r(\rho E + P)) = \rho V_r g_r. \quad (3)$$

Here, $E = \varepsilon + \frac{\mathbf{v}^2}{2}$ is the sum of the specific inner $(\varepsilon)$ and kinetic energies. The radial acceleration of gravity $g_r$ is the sum of two components: $g_r = g_{\rm pns} + g_{\rm env}$. The first component is attributable to the gravitational field produced by the PNS located exactly at the coordinate origin:

$$\mathbf{g}_{\rm pns} = \left(-\frac{GM_{\rm pns}}{r^2}, 0, 0\right). \quad (4)$$

The second component is attributable to the intrinsic gravitational field of the stellar envelope. The acceleration of this field is defined by the standard equation $\mathbf{g}_{\rm env} = -\nabla\Phi$ and the potential satisfies the Poisson equation

$$\Delta\Phi = 4\pi G\rho. \quad (5)$$

In our calculations, we actually used a numerical scheme based on the equations of ideal hydrodynamics written in axisymmetric form $\left(\frac{\partial}{\partial\varphi},\ g_\varphi = 0\right)$ in spherical coordinates $(r, \theta, \varphi)$:

$$\frac{\partial \rho}{\partial t} + \frac{1}{r^2}\frac{\partial}{\partial r}(r^2 \rho V_r) + \frac{1}{r\sin\theta}\frac{\partial}{\partial\theta}(\sin\theta \rho V_\theta) = 0, \quad (6)$$

$$\frac{\partial \rho V_r}{\partial t} + \frac{1}{r^2}\frac{\partial}{\partial r}(r^2 \rho V_r^2) + \frac{1}{r\sin\theta}\frac{\partial}{\partial\theta}(\sin\theta \rho V_r V_\theta) + \frac{\partial P}{\partial r} - \frac{\rho(V_\theta^2 + V_\varphi^2)}{r} = \rho g_r, \quad (7)$$

$$\frac{\partial \rho V_\theta}{\partial t} + \frac{1}{r^2}\frac{\partial}{\partial r}(r^2 \rho V_r V_\theta) + \frac{1}{r\sin\theta}\frac{\partial}{\partial\theta}(\sin\theta \rho V_\theta^2) + \frac{1}{r}\frac{\partial P}{\partial\theta} + \frac{\rho V_r V_\theta}{r} - \frac{\rho V_\varphi^2 \cot\theta}{r} = \rho g_\theta, \quad (8)$$

$$\frac{\partial \rho V_\varphi}{\partial t} + \frac{1}{r^2}\frac{\partial}{\partial r}(r^2 \rho V_r V_\varphi) + \frac{1}{r\sin\theta}\frac{\partial}{\partial\theta}(\sin\theta \rho V_\theta V_\varphi) + \frac{\rho V_\varphi}{r}(V_r + \cot\theta V_\theta) = 0, \quad (9)$$

$$\frac{\partial \rho E}{\partial t} + \frac{1}{r^2}\frac{\partial}{\partial r}(r^2 V_r(\rho E + P)) + \frac{1}{r\sin\theta}\frac{\partial}{\partial\theta}(\sin\theta V_\theta(\rho E + P)) = \rho(V_r g_r + V_\theta g_\theta). \quad (10)$$

As can be easily verified, this system transforms to the system of equations (1)–(3) if we set $V_\theta$, $V_\varphi$, and $g_\theta$ equal to zero.

Equations (6)–(10) are written in divergence form; since they do not contain viscosity, including artificial viscosity, shock and contact discontinuities can emerge in their solutions. The absence of heat conductivity in these equations makes discontinuities in specific internal energy and density admissible. This can be taken into account when choosing a finite-difference scheme.

### *Initial Data*

As we noted in the Introduction, the formation time of the PNS embryo with the mass ($\sim 1 M_\odot$) characteristic of our problem, when the neutron signal also increases most steeply (even before its maximum), may by arbitrarily taken as the initial time ($t = 0$). In that case, the outer layers may still be considered in a hydrostatic equilibrium. We used the distributions of thermodynamic quantities obtained in the studies of the evolution of massive stars (Boyes *et al.* 1999) as the initial data. From this paper, we took the density and temperature profiles, which, in turn, serve to determine the initial pressure and internal energy by using a specified equation of state. Of course, the equation of state used in our calculations slightly differs from the equation of state used in obtaining these profiles. The main difference is that we took a large number of isotopes into account in our numerical calculations of stellar evolution. However, a detailed comparison of the equations of state shows that, quantitatively, these differences are moderately large and passing to a simpler equation of state in our calculations has no significant effect on the radial distributions of thermodynamic quantities. The pressure calculated from fixed density and temperature using our equation of state differs by no more than 2% along the entire profile from the pressure taken from Boyes *et al.* (1999); the pressure from Boyes *et al.* (1999) is lower approximately by 2% near the iron core and higher by the same ∼2% in the profile tail. Nevertheless, to exactly specify hydrostatically equilibrium distributions of thermodynamic quantities in our calculations as the initial data, we recalculated the initial profiles using a new equation of state. As was noted above, by the formation of a PNS embryo with a typical mass of $\sim 1M_\odot$, the outer layers surrounding the iron core are virtually in hydrostatic equilibrium, as suggested by hydrodynamic collapse calculations (Nadyozhin 1977). To be more precise, the radial temperature distribution from Boyes *et al.* (1999) was used in these calculations and the new distributions of pressure and the remaining thermodynamic quantities $(\rho, e)$ were reconstructed by solving the system of hydrostatic equilibrium equations

$$\frac{\partial P}{\partial r} = -\frac{Gm\rho}{r^2}, \quad (11)$$

$$\frac{\partial m}{\partial r} = 4\pi r^2 \rho; \quad (12)$$

the pressure $P$ and mass $m$ at the inner boundary were also taken from Boyes *et al.* (1999). The dimensions of the computed region were chosen in such a way that the inner boundary was equal to the radius that bounded the region of mass $1M_\odot$ in the initial profile and the outer (in radius) boundary coincided with the inner boundary of the helium shell. The calculations were performed for the evolutionary models of stars with masses of 11, 20, and $25M_\odot$ and solar metallicity $Z$.

The initial values of the velocity component $V_r$ were taken to be zero in all cases, except for an additional calculation in which we imparted a velocity approximately equal to the free-fall velocity in magnitude and directed away from the center.

In all of our calculations, we took into account the presence of a neutrino signal. The gravitational collapse of massive stellar cores gives rise to a powerful neutrino signal, as predicted by Nadyozhin (1978) and experimentally confirmed by the observations of the SN 1987A explosion in the Large Magellanic Cloud. The neutrino signal carries away almost all of the gravitational binding energy of the forming neutron star, which is released during the collapse. The total gravitational mass defect of the neutron star was calculated by using the formula $\Delta M_G = 0.084\frac{M_{\rm Fe}^2}{M_\odot}$ (Lattimer and Yahil 1989), which is $\Delta M_G = 0.054$, 0.165, and $0.336M_\odot$ for iron cores with masses $M_{\rm Fe} = 0.8$, 1.4, and $2.0M_\odot$, respectively. It should be immediately noted that these values of $\Delta M_G$ refer to cold neutron stars; i.e., they definitely overestimate the effect under consideration, because a hot PNS still has high internal energy and temperature. The shape of the neutrino-signal curve in time was taken from Nadyozhin (1978). During the calculation, it was fitted by a set of exponentials that were continuously joined at the boundary points.

*Boundary Conditions*

The region of the solution of the problem or the computed region has the shape of a spherical envelope (see above), $r_{\rm min} \le r \le r_{\rm max}$; the choice of $r_{\rm min}$ and $r_{\rm max}$ is determined by the physical considerations outlined in the Subsection "Initial Data." Sufficient boundary conditions must be specified precisely on these constant values of the Eulerian radius.

Thus, in our problem, the computed region has the Eulerian outer and inner boundaries, $r_{\rm min}$ and $r_{\rm max}$. The inner boundary is a transparent wall; i.e., the derivatives of all physical quantities ($v_r, \rho$, and $\varepsilon$) are assumed to be zero. Matter can then freely flow through the inner boundary. In that case, the accretion rate or, more precisely, the total mass of the matter passing through it per unit time is calculated at this boundary. At the outer boundary, the condition simulates a vacuum outside the computed region; i.e., the thermodynamic quantities ($\rho, \varepsilon$, and $P$) are set equal to nearly zero. The boundary condition for the gravitational potential is $\Phi \to -GM/r$ for $r \to \infty$. Recall that we use the numerical scheme of a two-dimensional problem (see the Section "The System of Equations").

## THE EQUATION OF STATE

We use the equation of state for matter treated as a mixture of a perfect Boltzmann gas of nuclei, a perfect Fermi–Dirac electron–positron gas, and blackbody radiation. In a wide temperature range, the matter is assumed to be a mixture of a perfect iron gas of nuclei, a perfect electron–positron gas, and blackbody radiation. However, in order that the equation of state be applicable under nuclear-statistical-equilibrium (NSE) conditions, i.e., at high temperatures $T > T_{\rm c}$, where $T_{\rm c} = (3–5) \times 10^9$ K (Imshennik and Nadyozhin 1965, 1982), virtually independent of the matter density $\rho$ (Imshennik *et al.* 1981), we assume that at $T > T_{\rm c}$, the matter is a mixture of perfect gases of four nuclides, ${}^1_0n$, ${}^1_1p$, ${}^4_2$He, ${}^{56}_{26}$Fe, with a perfect gas of $e^-$, $e^+$ leptons and blackbody radiation. Including many other nuclides should not significantly affect the thermodynamic functions of the matter.

The contribution of the electron–positron component is given by general integral equations with arbitrary degeneracy and relativity. The difference between the electron and positron number densities can be determined from the necessary condition of electrical neutrality. In turn, the latter in the form of a simple integral equation serves to calculate the electron chemical potential ($\mu$):

$$\frac{1}{\lambda^3}\int_0^\infty [F_-(\xi) - F_+(\xi)]d\xi \tag{13}$$
$$= \frac{\rho}{m_u}\left(\frac{26}{56}X_{\rm Fe} + \frac{1}{2}X_{\rm He} + X_p\right),$$

where $X_{\rm Fe}$, $X_{\rm He}$, and $X_p$ are the mass fraction of iron, helium, and free protons; and $F_\pm(\xi)$ under the integral are the Fermi–Dirac functions. For arbitrary degeneracy and relativity of $e^\pm$ leptons, these functions can be expressed explicitly:

$$F_\pm(\xi) = \xi^2\left(1 + \exp z_\pm(\xi)\right)^{-1}, \tag{14}$$
$$z_\pm(\xi) = \alpha(\sqrt{1+\xi^2} \pm w),$$

because $\mu \equiv \mu_- = -\mu_+$. Here, we introduced the dimensionless parameters $\alpha \equiv \dfrac{m_{\rm e}c^2}{k_{\rm B}T}$ and $w \equiv \dfrac{\mu}{m_{\rm e}c^2}$, where $T$ is the temperature, $k_{\rm B}$ is the Boltzmann constant, $m_{\rm e}$ is the electron mass, $\mu$ is the chemical potential (including the fermion rest energy), and the quantity $\lambda$ as the natural unit of length (of the order of the Compton length) $\lambda^3 = \dfrac{1}{8\pi}\left(\dfrac{h}{m_{\rm e}c}\right)^3$.

The equation of state proper, i.e., the expressions for the pressure and specific internal energy of the matter, is then (Imshennik and Nadyozhin 1965, 1982)

$$P = P_- + P_+ + \frac{1}{3}aT^3 \tag{15}$$
$$+ \frac{k_{\rm B}}{m_u}\rho T\left(X_n + X_p + \frac{1}{4}X_{\rm He} + \frac{1}{56}X_{\rm Fe}\right),$$

$$\varepsilon = \varepsilon_- + \varepsilon_+ + \frac{aT^4}{\rho} + \frac{3}{2}\frac{k_{\rm B}}{m_u}T \tag{16}$$
$$\times\left(X_n + X_p + \frac{1}{4}X_{\rm He} + \frac{1}{56}X_{\rm Fe}\right)$$
$$+\left[\frac{Q_{\rm Fe} + 26\Delta Q_n}{56 m_u}(1 - X_{\rm Fe})\right.$$
$$\left. - \frac{Q_{\rm He} + 2\Delta Q_n}{4m_u}X_{\rm He} - \frac{\Delta Q_n}{m_u}X_p\right].$$

Here, $\Delta Q_n = (m_n - m_p)c^2 = 1.294$ MeV is the energy threshold of $\beta$ decay and $a = 7.5644 \times 10^{-15}$ erg cm$^{-3}$ K$^{-1}$ is the neutron radiation density constant.

The contributions of the lepton pressure and specific internal energy to the complete equation of state are given by standard general relations (Landau and Lifshitz 1976) [with definitions (14)]:

$$P_- + P_+ = \frac{m_{\rm e}c^2}{3\lambda^3}\int_0^\infty \frac{\xi^2}{\sqrt{1+\xi^2}}[F_-(\xi) + F_+(\xi)]d\xi, \tag{17}$$

$$\varepsilon_- + \varepsilon_+ = \frac{m_{\rm e}c^2}{\lambda^3\rho}\int_0^\infty \sqrt{1+\xi^2}\,[F_-(\xi) + F_+(\xi)]\,d\xi \tag{18}$$
$$- m_{\rm e}c^2\left(\frac{26}{56}\frac{1}{m_u}\right).$$

Note that a constant electron rest energy at zero temperature (in the absence of positrons) is subtracted from the sum of $\varepsilon_+$ and $\varepsilon_-$. Clearly, the specific internal energies can generally be determined to within an arbitrary constant.

The mass fractions of iron, helium, free neutrons, and free protons under NSE conditions at a given density and temperature are defined by the system of equations (Imshennik and Nadyozhin 1965, 1982; Imshennik and Zabrodina 1999)

$$X_i = \omega_i A^{5/2}\left(\frac{h^2}{2\pi m_u k_{\rm B}T}\right)^{\frac{3}{2}(A_i-1)}\frac{1}{2^{A_i}} \tag{19}$$
$$\times\left(\frac{\rho}{m_u}\right)^{A_i-1} X_n^{A_i-Z_i}X_p^{Z_i}\exp\left(\frac{Q_i}{kT}\right),$$

$$\theta_0 = \frac{X_n + (1/2)X_{\rm He} + (30/56)X_{\rm Fe}}{X_p + (1/2)X_{\rm He} + (26/56)X_{\rm Fe}} = \frac{30}{26}, \tag{20}$$

$$X_n + X_p + X_{\rm He} + X_{\rm Fe} = 1, \tag{21}$$

where $i = $ He, Fe; $m_u$ is the atomic mass unit; $Q_i$ is the binding energy in the ground state of the nuclides ($Q_{\rm He} = 28.296$ MeV, $Q_{\rm Fe} = 492.26$ MeV); $A_i$ and $Z_i$ are the mass and charge numbers of these nuclides; and $\omega_i$ are the partition functions, which may be taken to be unity without any particular error. Here, we assumed the ratio of the total numbers of neutrons and protons to be constant; $\theta = 30/26 = \theta_{\rm Fe}$ is the value typical for the $^{56}_{26}$Fe nuclide (Imshennik and Zabrodina 1999). Actually, this implies that we completely ignore the $\beta$ processes and disregard the variety of iron-peak elements among the explosion products of a low-mass neutron star and in the surrounding iron gas. Clearly, only iron nuclei are available at $T < T_{\rm c}$; i.e., $X_{\rm Fe} = 1$ and $X_{\rm He} = X_n = X_p = 0$.

The term in the square brackets from (16) is the rest energy of the matter as a function of the mass fractions, which, obviously, becomes zero at $X_{\rm Fe} = 1$. This means that the energy of pure iron is taken as a zero level, being, of course, the lowest value of the above term.

Since the pressure and the specific internal energy must regularly be used in solving the system of hydrodynamic equations, it would be appropriate to use a tabulated equation of state. This would allow us to avoid complex calculations at each step of the solution of system (6)–(10) if the thermodynamic functions were calculated once and with a high accuracy. Our numerical method for solving the system of hydrodynamic equations (6)–(10) uses the density and specific internal energy as the main variables. Therefore, the equation of state is $P = f(\rho, \varepsilon)$. In the numerical method (see below), this equation is simulated by the so-called binomial equation of state

$$P = \left[(\bar{\gamma} - 1)\,\varepsilon + c_0^2\right]\rho - \rho_0 c_0^2. \tag{22}$$

This simulation is local in nature. The constants $\bar{\gamma}$, $c_0^2$, and $\rho_0$ are determined by the pressure and its derivatives at a given point:

$$c_0^2 = \left(\frac{\partial f}{\partial \rho}\right)_\varepsilon - \frac{\varepsilon}{\rho}\left(\frac{\partial f}{\partial \varepsilon}\right)_\rho, \tag{23}$$

$$\rho_0 c_0^2 = \rho\left(\frac{\partial f}{\partial \rho}\right)_\varepsilon - f, \quad \bar{\gamma} - 1 = \frac{1}{\rho}\left(\frac{\partial f}{\partial \varepsilon}\right)_\rho.$$

By directly substituting (23) into (22), it can be verified that the simulation of equation of state (22) with specified [according to (23)] constants satisfies the requirement that the pressure and the speed of sound derived from (22) be equal to those derived from the real equation of state $P = f(\rho, \varepsilon)$. Using the binomial approximation for the equation of state considerably simplifies the solution of the problem of discontinuity breakup, which underlies the numerical method. Thus, for the system of hydrodynamic equations to be numerically solved, we must know the pressure and its first derivatives with respect to density and specific internal energy as a function of these variables.

Equations (15)–(16) do not allow us to obtain $P$ as a function of $\rho$ and $\varepsilon$ in explicit form. Therefore, it is necessary to first determine all of the thermodynamic quantities concerned as functions of density and temperature. To this end, we used an approach related to the Gaussian quadrature method (Krylov 1967). This approach has a sufficient accuracy over the entire range of densities and temperatures concerned (Blinnikov *et al.* 1996). To solve the system of hydrodynamic equations (6)–(10), we must also know the first derivatives of the pressure with respect to density $\rho$ and specific internal energy $\varepsilon$. Using standard relations for the derivatives of thermodynamic quantities (Landau and Lifshitz 1976), we find the thermodynamic derivatives of the pressure needed for the numerical scheme to be implemented

$$\left(\frac{\partial P}{\partial \rho}\right)_\varepsilon = \left(\frac{\partial \tilde{P}}{\partial \rho}\right)_T - \frac{\left(\frac{\partial \tilde{P}}{\partial T}\right)_\rho}{\left(\frac{\partial \varepsilon}{\partial T}\right)_\rho}\left(\frac{\partial \varepsilon}{\partial \rho}\right)_T, \tag{24}$$

$$\left(\frac{\partial P}{\partial \varepsilon}\right)_\rho = \frac{\left(\frac{\partial \tilde{P}}{\partial T}\right)_\rho}{\left(\frac{\partial \varepsilon}{\partial T}\right)_\rho},$$

where $\tilde{P}(\rho, T) = P(\rho, \varepsilon(\rho, T))$.

Finally, for the equation of state to be derived in final form, we must reinterpolate the computed tables of thermodynamic quantities and their derivatives, which depend on density and temperature, to the domain of the $\rho$ and $\varepsilon$ variations.

Clearly, using our equation of state, we can simulate the formation of a nucleon bubble within the computed region by artificially raising the temperature in the initial conditions compared to the temperature obtained by Boyes *et al.* (1999). In the above recalculations of the hydrostatic equilibrium conditions with the inevitable decrease in density on the simulation segments, we need not be concerned about the conservation of a purely nucleon composition, because the decrease in density only facilitates it.

## THE NUMERICAL METHOD OF SOLUTION

The numerical solution of our problem is based on the popular and universal PPM method. This method uses the Eulerian finite-difference scheme (Colella and Woodward 1984) and is a modification of Godunov's method.

The numerical calculation was performed in spherical coordinates. To construct the finite-difference scheme, we rewrote the initial system of equations (6)–(10) in variables $\mathbf{U} = \{\rho, \rho V_r, \rho V_\theta, \rho V_\varphi, \rho E\}^T$:

$$\frac{\partial \mathbf{U}}{\partial \mathrm{t}} + \frac{\partial\left(A_r \mathbf{F}_r(\mathbf{U})\right)}{\partial D_r} + \frac{\partial\left(A_\theta \mathbf{F}_\theta(\mathbf{U})\right)}{\partial D_\theta} + \frac{\partial \mathbf{G}}{\partial r} + \frac{\partial \mathbf{H}}{\partial \theta} = \mathbf{J},$$

$$\mathbf{F}_r(\mathbf{U}) = \begin{pmatrix} \rho V_r \\ \rho V_r^2 \\ \rho V_r V_\theta \\ \rho V_r V_\varphi \\ \rho V_r E + V_r P \end{pmatrix},$$

$$\mathbf{F}_\theta(\mathbf{U}) = \begin{pmatrix} \rho V_\theta \\ \rho V_\theta V_r \\ \rho V_\theta^2 \\ \rho V_\theta V_\varphi \\ \rho V_\theta E + V_\theta P \end{pmatrix}, \quad (25)$$

$$\mathbf{G} = \begin{pmatrix} 0 \\ P \\ 0 \\ 0 \\ 0 \end{pmatrix}, \quad \mathbf{H} = \begin{pmatrix} 0 \\ 0 \\ P/r \\ 0 \\ 0 \end{pmatrix},$$

$$\mathbf{J} = \begin{pmatrix} 0 \\ \rho g_r + \rho(V_\theta^2 + V_\varphi^2)/r \\ \rho g_\theta + \rho(V_\varphi^2 \cot\theta - V_r V_\theta)/r \\ -\rho V_\varphi (V_r + V_\theta \cot\theta)/r \\ \rho(g_r V_r + g_\theta V_\theta) \end{pmatrix},$$

$$A_r = r^2, \quad D_r = \frac{r^3}{3},$$

$$A_\theta = \frac{\sin\theta}{r}, \quad D_\theta = -\cos\theta.$$

System (25) is hyperbolic; to solve it, we broke down the computed region into cells using a grid composed of surfaces of constant radius ($r$) and constant polar angle ($\theta$). Next, we formally averaged the equations of system (25) over the cell volume $\Delta V_{ijk}$ and over the time step $\Delta t = t^{n+1} - t^n$. The problem of determining the distribution of the physical quantities $\{\rho, \rho V_r, \rho V_\theta, \rho V_\varphi, \rho E\}_{ij}$ at a later time $t^{n+1}$ from their known distribution at a given time $t^n$ is then reduced to determining the time-averaged fluxes at the cell boundaries and the time- and cell-volume-averaged free terms if, of course, the latter exist. To calculate the average fluxes, the interaction between two adjacent cells is considered as the problem of discontinuity breakup, whose solution is the main procedure of our numerical method. In general, an arbitrary discontinuity self-similarly breaks up into a configuration composed of four flow regions, which differ in properties and which are separated from one another by shock (or rarefaction) waves and a contact discontinuity (see, e.g., Landau and Lifshitz 1986; Kibel *et al.* 1963). In constructing the finite-difference scheme, we simulated the equation of state by the binomial formula (22). The use of this formula made it possible to properly describe strong rarefaction waves, which is of particular importance in astrophysical problems characterized by density variations over a wide range and by large density jumps. In this case, we can avoid numerically solving a system of ordinary differential equations and, without resorting to the formal replacement of rarefaction waves by shock waves (Colella and Glas 1985), can easily write out analytic formulas for shock and rarefaction waves. These formulas are required to construct an iterative process of calculating the pressure and velocity in the central region formed after a discontinuity breakup.

Godunov's method can be generalized to a multidimensional case (Godunov *et al.* 1976). However, the difficulty in determining the regions of influence for the problem of discontinuity breakup arises for our algorithm based on the Eulerian scheme. A possible solution of this problem is that several regions of influence, each corresponding to a certain characteristic (Riemann invariant), are found for the boundary between adjacent cells. The interpolation distributions of the physical quantities are averaged over all the derived regions of influence and the initial data for the problem of discontinuity breakup are obtained as the sums of these means with weighting coefficients proportional to the absolute values of the increments of the Riemann invariants along the corresponding directions. As a result, the characteristic properties of the system are properly taken into account (Dai and Woodward 1997). We use a parabolic interpolation for the physical quantities within a cell, which is monotonic and continuous at the cell boundaries in the region of smooth flow and conserves the total volume integrals for the quantities being interpolated (Colella and Woodward 1984). The constructed scheme is explicit; there is a restriction on the time step, the Courant condition, to maintain the stability of the calculation. In addition, the method does not require a special choice of artificial viscosity for each specific calculation. A finite-difference discrete approximation of the functions that describe the fields of the physical quantities automatically smears and smoothes out the discontinuities.

Below, we also give the system of units used in our calculations. The scales of the physical quantities in this system are

$$[r] = R_0, \quad [V_r] = [V_\theta] = [V_\varphi] = (GM_0/R_0)^{1/2}, \quad (26)$$

$$[\rho] = M_0/(4\pi R_0^3), \quad [t] = R_0^{3/2}/(GM_0)^{1/2},$$

$$[P] = GM_0^2/(4\pi R_0^4),$$

$$[E] = GM_0/R_0, \ [T] = (GM_0^2/(4\pi R_0^4 a_r))^{1/4},$$

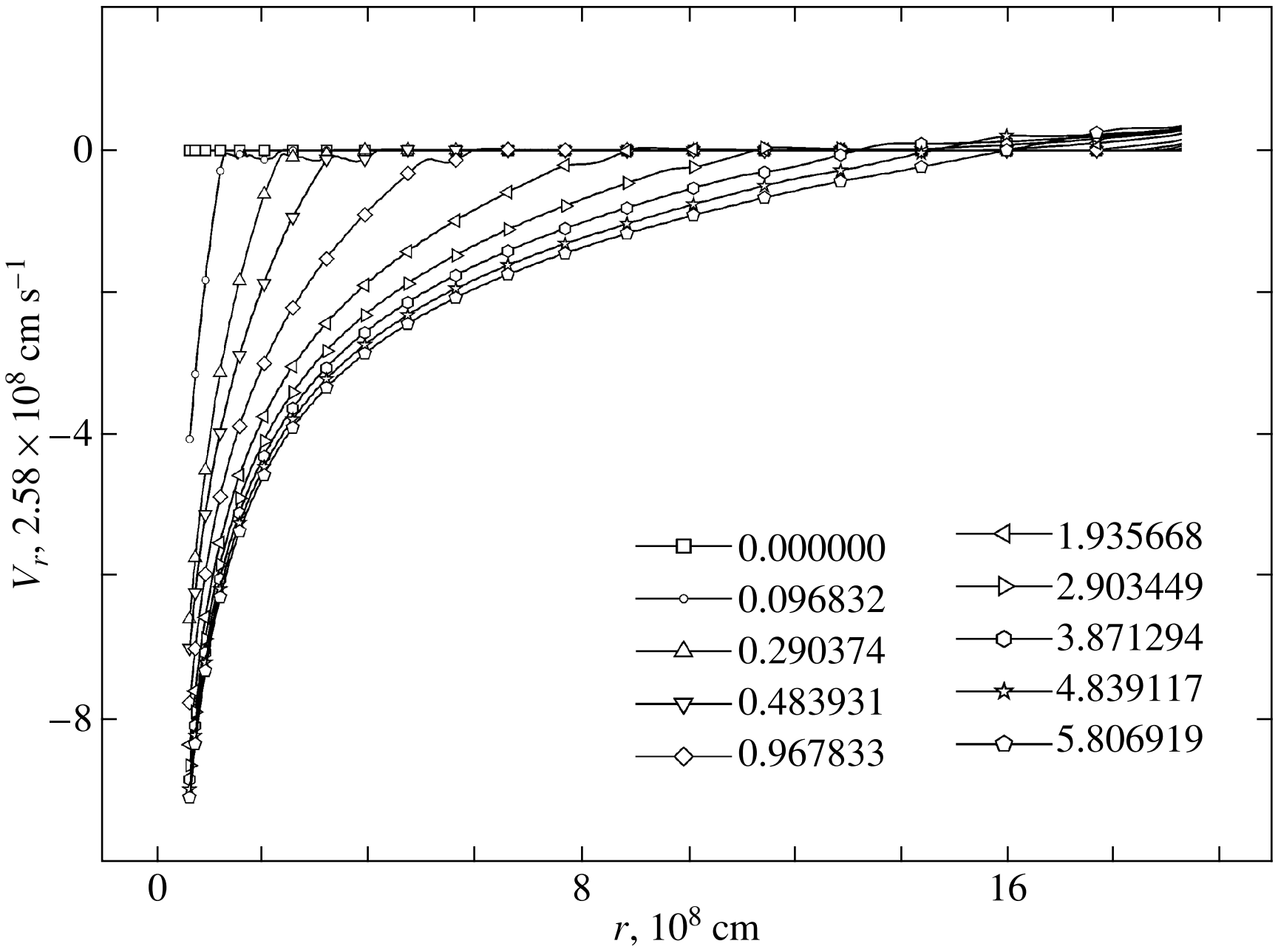

**Fig. 1.** The radial velocity component versus the radius at consecutive times (seconds) for data set [1].

where $R_0$ and $M_0$ are some length and mass scales. In units (26), the hydrodynamic equations with Newtonian gravitation contain no dimensionless parameters. It would be natural to use the following characteristic values from our problem as $R_0$ and $M_0$: $R_0 = 10^8$ cm and $M_0 = 10^{32}$ g. The numerical values for the scales of the physical quantities from (26) are then

$$[r] = 10^8 \text{ cm}, \tag{27}$$
$$[V_r] = [V_\theta] = [V_\varphi] = 2.583 \times 10^8 \text{ cm s}^{-1},$$
$$[\rho] = 7.958 \times 10^6 \text{ g cm}^{-3}, \quad [t] = 3.871 \times 10^{-1} \text{ s},$$
$$[P] = 5.310 \times 10^{23} \text{ erg cm}^{-3},$$
$$[E] = 6.674 \times 10^{16} \text{ erg}, \quad [T] = 2.894 \times 10^9 \text{ K}.$$

An efficient algorithm is used to solve the Poisson equation (5) and to determine the gravitational acceleration **g** . This algorithm is convenient to use in the finite-difference scheme for integrating the hydrodynamic equations on a stationary grid in spherical coordinates (Aksenov 1999). The method is based on the expansion of the potential in integral representation,

$$\Phi = -G \int \frac{\rho(\mathbf{r}', t)}{|\mathbf{r}' - \mathbf{r}|} d\mathbf{r}', \tag{28}$$

by defining the Legendre polynomials in terms of the generating function. Using the addition theorem for associated Legendre polynomials, expression (28) can be reduced to a form convenient for averaging over the cell volume in spherical coordinates. This procedure can be used in problems with any number of measurements and it is efficient: it requires $\sim l_{\max}$ operations per cell, where $l_{\max}$ is the number of associated Legendre polynomials used in the calculation. In addition, the boundary condition $\Phi \to -GM/r$ for $r \to \infty$ is automatically satisfied for the potential.

## NUMERICAL RESULTS

Let us turn to the results obtained when numerically integrating the system of hydrodynamic equations (6)–(10) with the initial data and boundary conditions given above. We performed two series of calculations: in the first series, we only took into account the effect of the decrease in the central gravitational mass due to the emission of a neutrino signal; in the second series, apart from allowance for this effect, we also analyzed the influence of hot nucleon bubbles that are presumably formed in the PNS corona.

In the first series of our calculations, we considered the following models of presupernova stars: [1] the total stellar mass $M_{\text{tot}} = 11M_\odot$, the iron-core mass (the mass of the central stellar region in which the mass fraction of iron-peak nuclei is more than 80%) $M_{\text{core}} = 1.266M_\odot$, the gravitational iron-core mass defect $\Delta M_G = 2.676 \times 10^{32}$ g, the mass of the matter inside the computed region $M_{\text{in}} = 0.627M_\odot$, the inner radius of the computed region $r_{\min} = 6.250 \times 10^7$ cm, its outer radius $r_{\max} = 1.937 \times 10^9$ cm, and the mass of the central region (up to the radius $r_{\min}$) $M_{\text{center}} = 0.9987M_\odot$; [2] $M_{\text{tot}} = 20M_\odot$, $M_{\text{core}} = 1.485M_\odot$, $\Delta M_G = 3.552 \times 10^{32}$ g, $M_{\text{in}} =$ $2.825M_\odot$, $r_{\rm min} = 6.952 \times 10^7$ cm, $r_{\rm max} = 6.308 \times 10^9$ cm, and $M_{\rm center} = 0.9992M_\odot$; [3] $M_{\rm tot} = 25M_\odot$, $M_{\rm core} = 1.620M_\odot$, $\Delta M_G = 4.386 \times 10^{32}$ g, $M_{\rm in} = 5.458M_\odot$, $r_{\rm min} = 8.764 \times 10^7$ cm, $r_{\rm max} = 6.607 \times 10^9$ cm, and $M_{\rm center} = 1.0001M_\odot$. In all of these calculations, we took the constant, relatively low specific internal energy $\varepsilon_{\rm b} = 10^{16}$ erg g$^{-1}$ and density $\rho_{\rm b} = 500$ cm s$^{-3}$ as the boundary condition at the outer (in radius) boundary $r_{\rm max}$ and assumed the radial velocity to be zero. A grid of 200 computed zones in radius was used in the numerical integration. An expansion in Legendre polynomials up to the number $l_{\rm max} = 20$ was used to solve the Poisson equation.

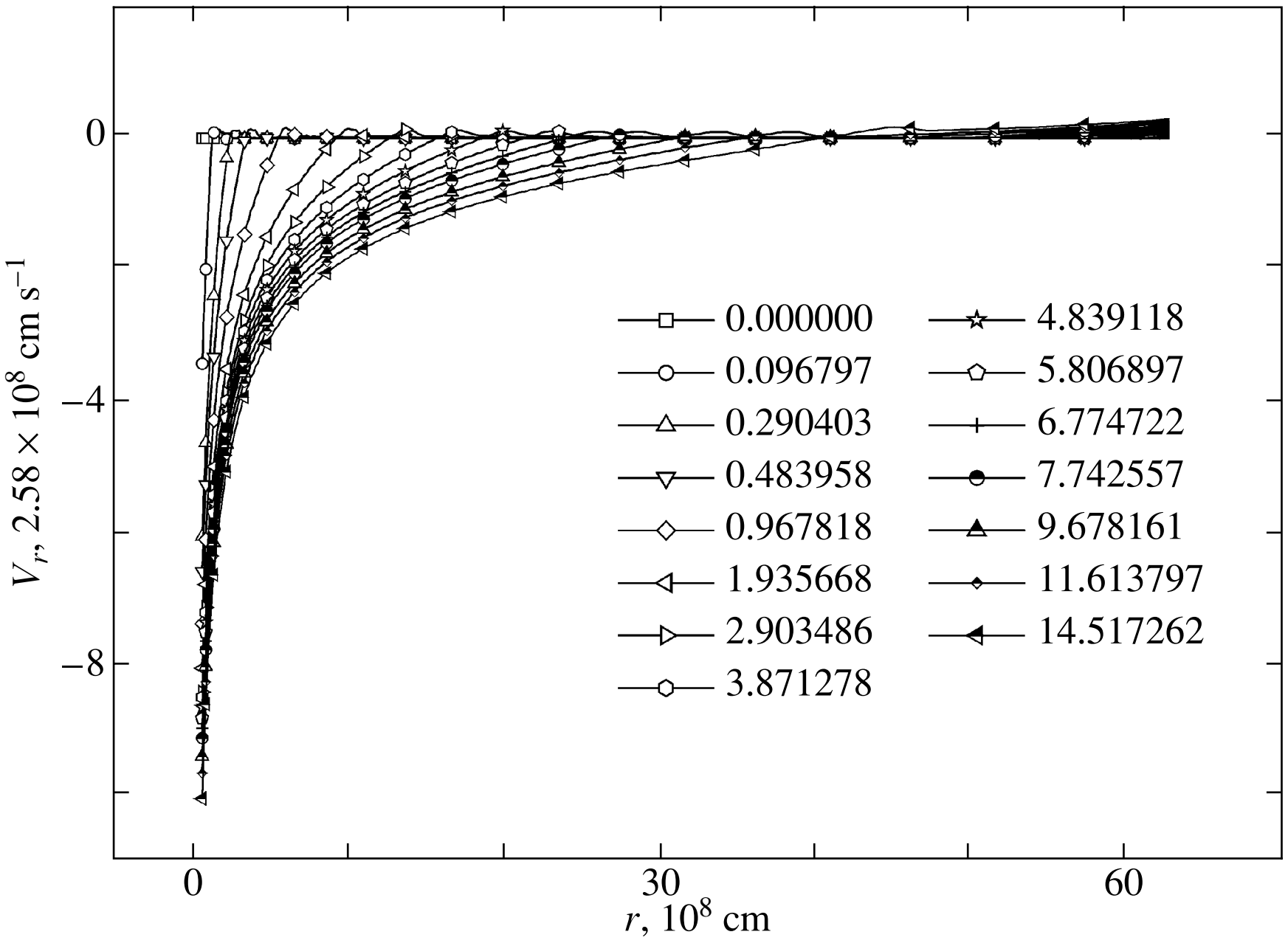

**Fig. 2.** Same as Fig. 1 for data set [2].

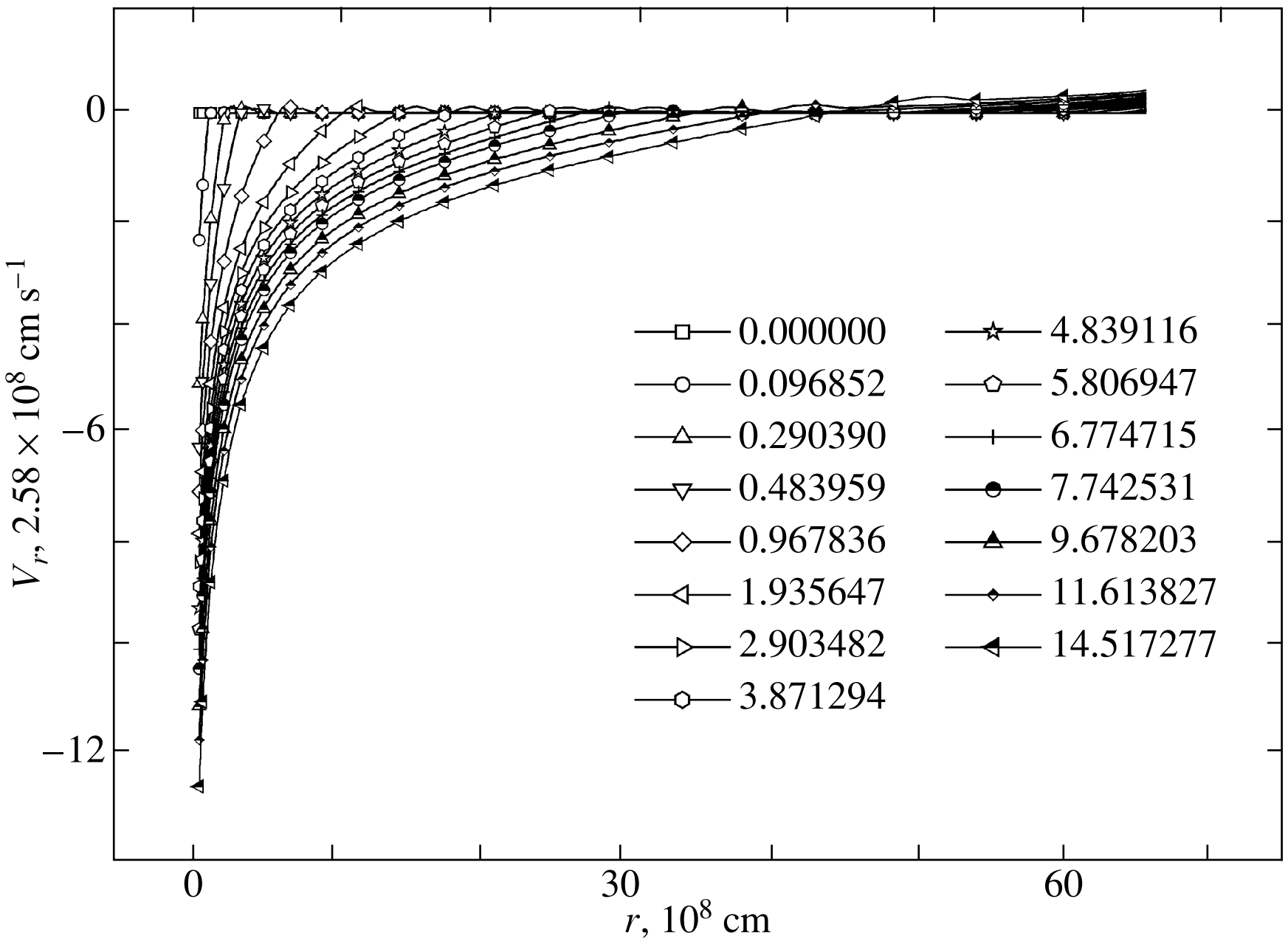

**Fig. 3.** Same as Fig. 1 for data set [3].

A similar flow pattern was observed in all of these cases. Figures 1–3 show plots of the radial veloc-ity against the radius (profiles) at consecutive times. As we see from these figures, the matter inside the computed region in all calculations falls through the transparent inner boundary to the center under the effect of gravity in a regime resembling free fall. The neutrino signal, which causes the mass of the PNS located at the center to decrease, manifests itself only in the existence of a modest maximum in the radial-velocity profile. In calculations [2, 3] with more massive stars ($20M_\odot$ and $25M_\odot$), the velocity maximum emerges immediately ahead of the front of a strong rarefaction wave: the maximum radial velocity is positive but low; it does not exceed $\sim 2.8 \times 10^7$ cm s$^{-1}$. In calculation [1] with a low-mass star ($11M_\odot$), the maximum lies in the range of negative velocities, so there are no positive radial velocities along the entire profile. We also see from Figs. 1–3 that in each case, a zone of positive velocities appears near the outer boundary of the computed region. This fact can be explained in terms of the boundary condition at the outer radius, which simulates a vacuum. This boundary condition has no fundamental effect on the properties of the flow as a whole; it only gives rise to a weak rarefaction wave. The main conclusion drawn from all calculations [1]–[3] is that almost all of the matter inside the computed region is accreted onto the central PNS. The plots of the radial velocity against the mass coordinate (profiles) show that the point of zero velocity (the point of separation) moves along the mass coordinate outward during the entire calculation. This implies that if the outer stellar envelope is actually ejected, then the point of separation cannot be closer to the center than the inner boundary of the helium shell.

The second series of our calculations was devoted to the effect of hot nucleon bubbles, which are most likely formed in the PNS corona, on the behavior of the onionlike structure of a massive star after the onset of gravitational iron-core collapse. In the initial data of these calculations, a hot bubble was specified in the following artificial way: the temperature in several adjacent cells near the inner boundary of the computed region was raised compared to the surrounding matter to the extent that, according to the adopted equation of state, the mass fractions of iron and helium nuclei became zero. The distribution of the remaining thermodynamic quantities was constructed to satisfy the hydrostatic equilibrium conditions (11), (12) as before. All calculations were performed with the same model of a $25M_\odot$ star; only the mass of the hot nucleon gas in the bubble was varied. Since the problem was solved in the one-dimensional approximation, the region of hot nucleon gas was actually a spherical shell rather than a bubble. The system of hydrodynamic equations (6)–(10) was integrated with the following initial data: [4] the minimum bubble radius $r_{\rm min}^{\rm NB} = 8.901 \times 10^7$ cm, the maximum bubble radius $r_{\rm max}^{\rm NB} = 1.042 \times 10^8$ cm, the hot-gas mass $M_{\rm NB} = 0.396 \times 10^{-2} M_\odot$, and the mass of the matter inside the computed region $M_{\rm in} = 5.779 M_\odot$; [5] $r_{\rm min}^{\rm NB} = 8.901 \times 10^7$ cm, $r_{\rm max}^{\rm NB} = 1.248 \times 10^8$ cm, $M_{\rm NB} = 0.693 \times 10^{-2} M_\odot$, $M_{\rm in} = 5.858 M_\odot$; [6] $r_{\rm min}^{\rm NB} = 8.901 \times 10^7$ cm, $r_{\rm max}^{\rm NB} = 1.499 \times 10^8$ cm, $M_{\rm NB} = 0.996 \times 10^{-2} M_\odot$, and $M_{\rm in} = 6.036 M_\odot$. Calculation [7] differs from case [6] only in that a positive velocity, $V_{\rm NB} = 1.734 \times 10^9$ cm s$^{-1}$, equal in magnitude to the local free-fall velocity was assigned to the matter in the cells that fell within the region of hot gas. In calculations [4]–[7], the parameters of the computed region were taken to be identical: the total stellar mass $M_{\rm tot} = 25M_\odot$, the iron-core mass $M_{\rm core} = 1.620 M_\odot$, the gravitational iron-core mass defect $\Delta M_G = 4.386 \times 10^{32}$ g, the inner radius of the computed region $r_{\rm min} = 8.764 \times 10^7$ cm, its outer radius $r_{\rm max} = 6.607 \times 10^9$ cm, and the mass of the central region $M_{{\rm center}r} = 1.0001 M_\odot$. Thus, all of the latter parameters are identical to those in case [3].

In Figs. 4–7, the radial velocity is plotted against radius at consecutive times for cases [4]–[7], respectively. As we see from these figures, the shock wave produced through energy release in the nucleon bubble due to proton and neutron recombination into iron nuclei has an amplitude that is qualitatively proportional to the initial mass of the hot gas. In case [4], the shock wave is rather weak and the velocity of the post-shock matter barely reaches $\sim 3 \times 10^8$ cm s$^{-1}$. In this case, a strong rarefaction wave is formed at the inner boundary of the computed region, as in the calculations that only took into account the gravitational effect of neutrino radiation, and the matter begins to be accreted onto the gravitating center at a high rate. We see from Fig. 8 that almost all of the matter that was initially inside the computed region passes through the inner boundary, whereas the escape of the matter through the outer boundary is negligible and is entirely attributable to the artificial rarefaction wave produced by a fixed boundary condition. The time it took for the shock wave to reach the helium shell is $\sim 13$ s. In cases [6]–[7], the nucleon-gas mass is at a maximum ($\sim 0.01 M_\odot$). The appearance of a strong shock wave and envelope ejection were shown in both cases: the post-shock radial velocity is $\sim 1.1 \times 10^9$ and $\sim 1.5 \times 10^9$ cm s$^{-1}$ for calculations [6] and [7], respectively. Therefore, the positive velocity ($\sim 1.7 \times 10^9$ cm s$^{-1}$) imparted to the hot gas of the nucleon bubble at the initial time in case [7] has a marginal effect on the envelope ejection. The rarefaction wave formed at the inner computed boundary is found to be much weaker than that in calculation [4]. As a result, the amount of matter accreted onto the center in these two calculations is modest: $\sim 0.045 M_\odot$ and $\sim 0.020 M_\odot$ for a zero and nonzero velocity of the hot gas, respectively. Almost all of the matter inside the computed region is ejected outward by the shock wave, as clearly confirmed by the plots in Figs. 9 and 10. In case [6], it took $\sim$6 s for the shock wave to reach the helium shell and $\sim$8 s for all of the matter to pass through the outer boundary; in case [7], both processes take about 4 s.

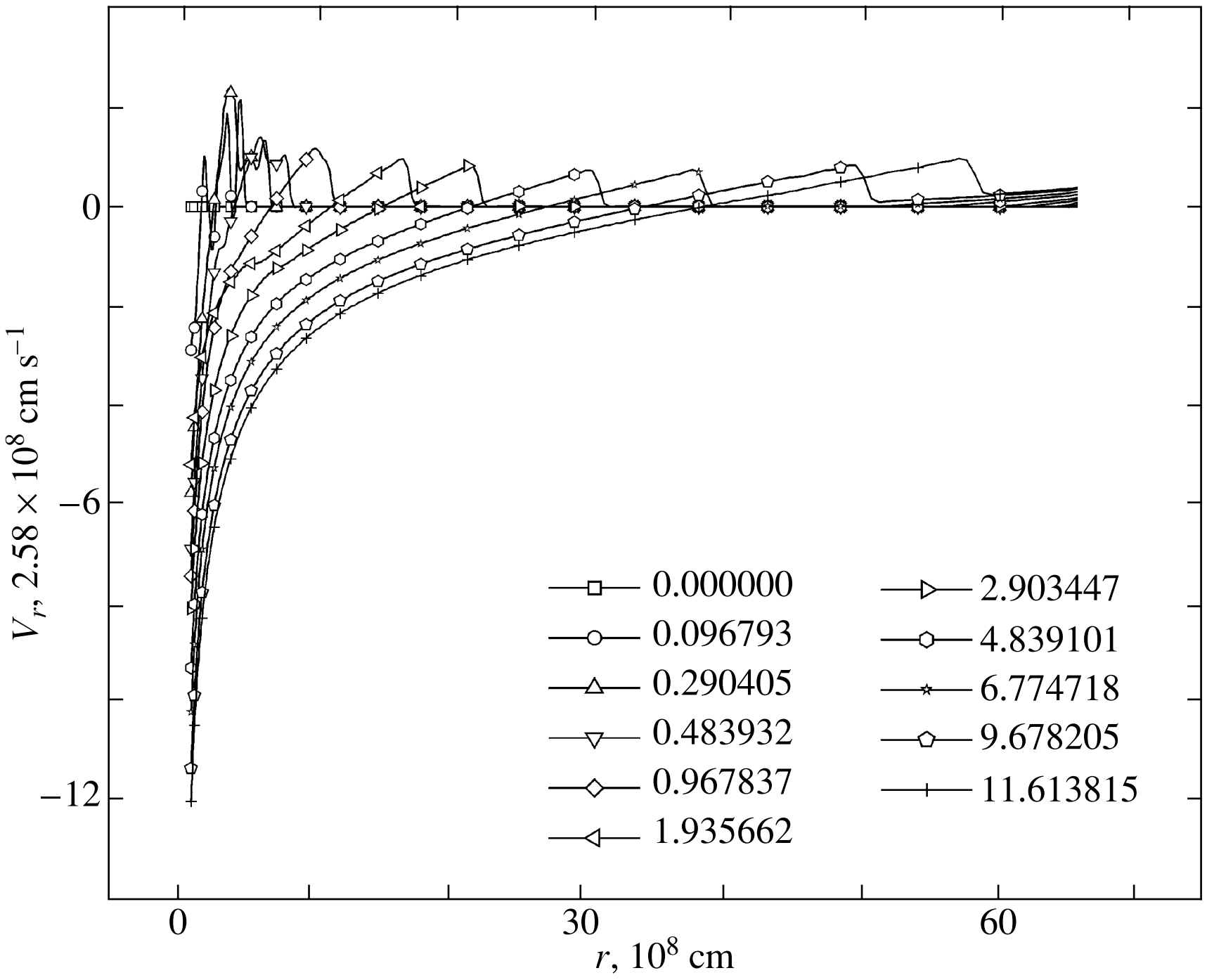

**Fig. 4.** Same as Fig. 1 for data set [4].

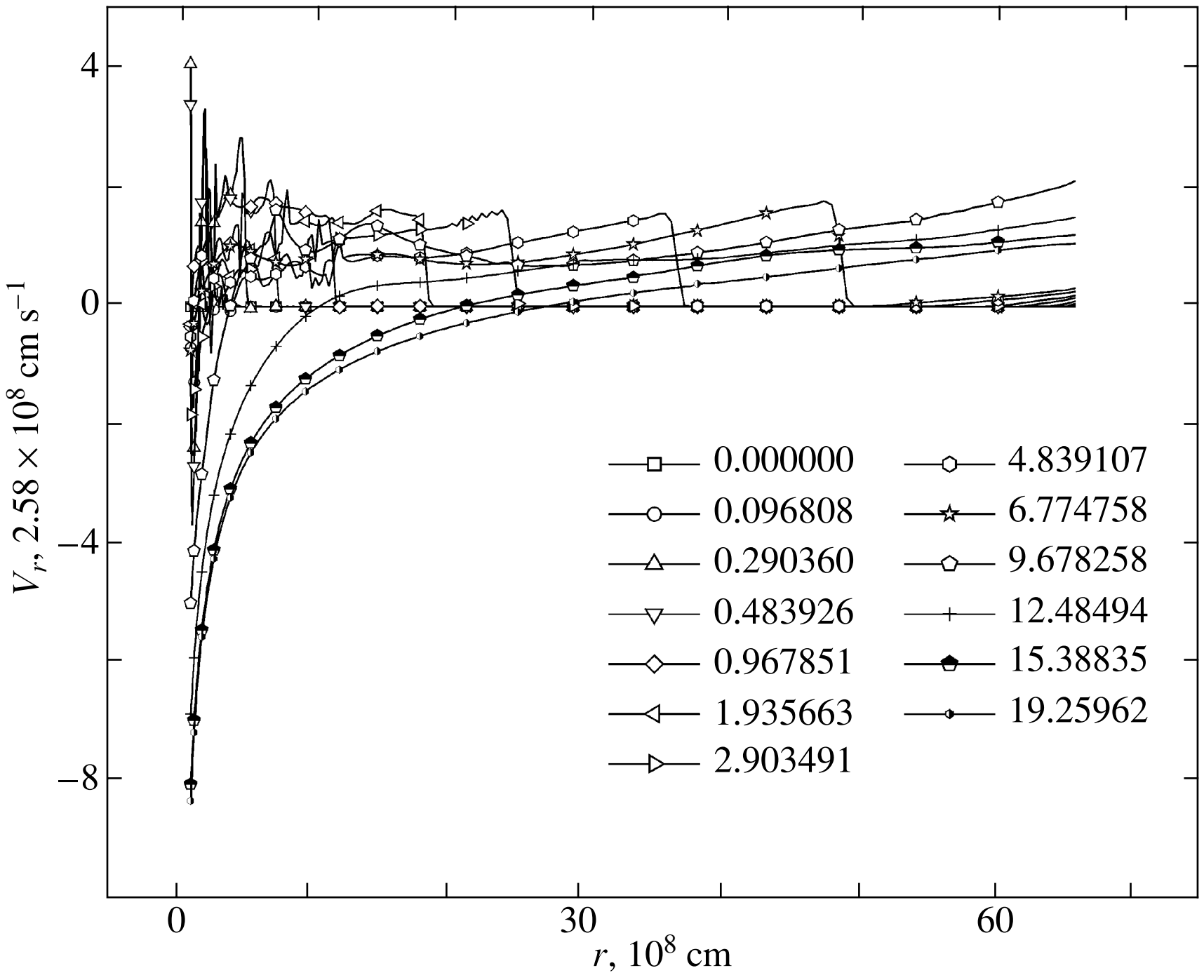

**Fig. 5.** Same as Fig. 1 for data set [5].

The calculation with the initial data [5] is interme-diate between case [4], on the one hand, and cases [6] and [7], on the other hand. A moderate rarefaction wave is formed near the inner boundary and the post-shock gas velocity is $\sim 5 \times 10^8$ cm s$^{-1}$ (Fig. 5). The shock wave reaches the helium shell in $\sim$10 s. In this case, mass accretion at the inner computed boundary begins with an appreciable delay of $\sim$12 s and the matter flows outward from the center at a rate that is approximately twice the rate of the gas accretion onto the PNS (Fig. 11).

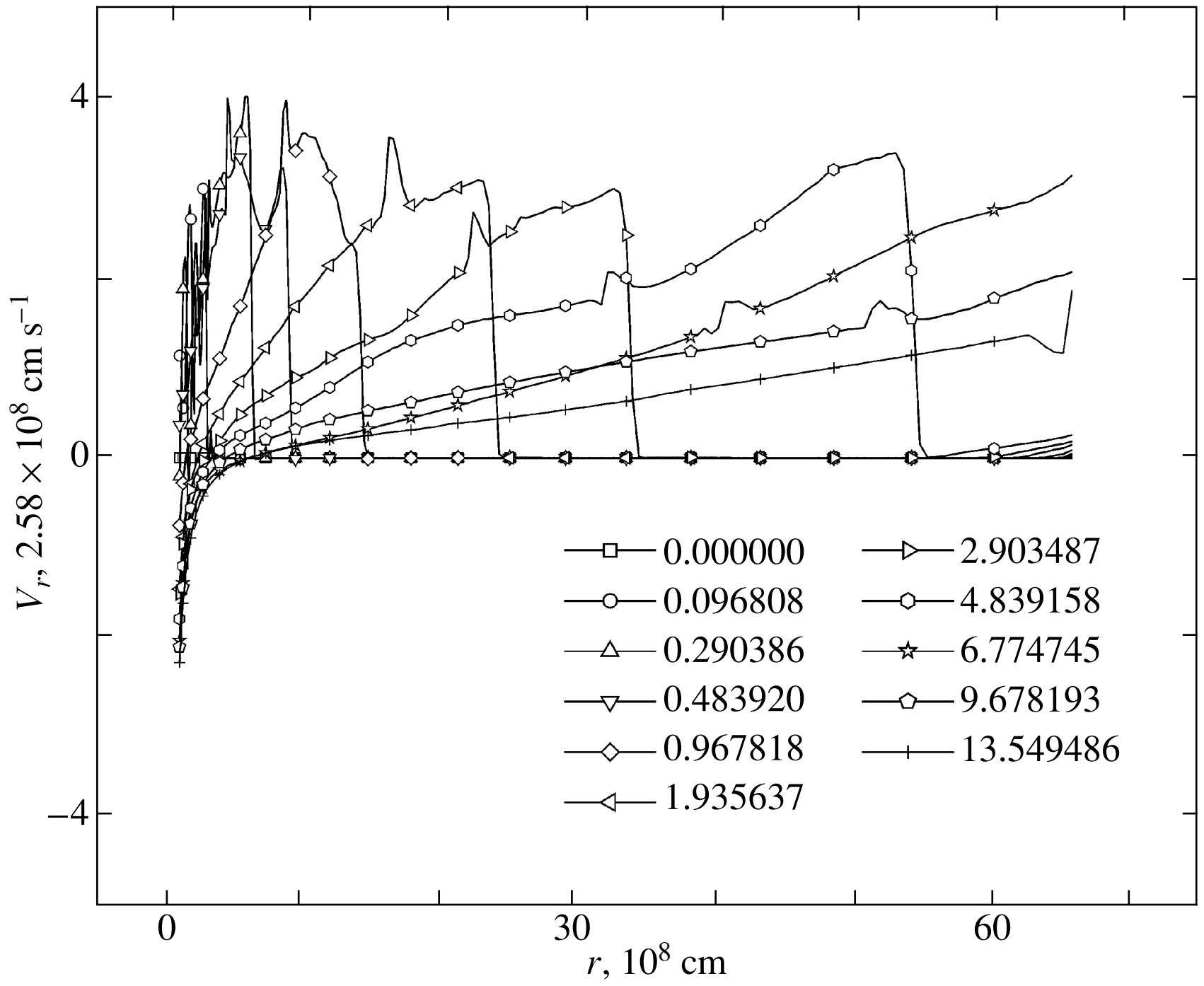

**Fig. 6.** Same as Fig. 1 for data set [6].

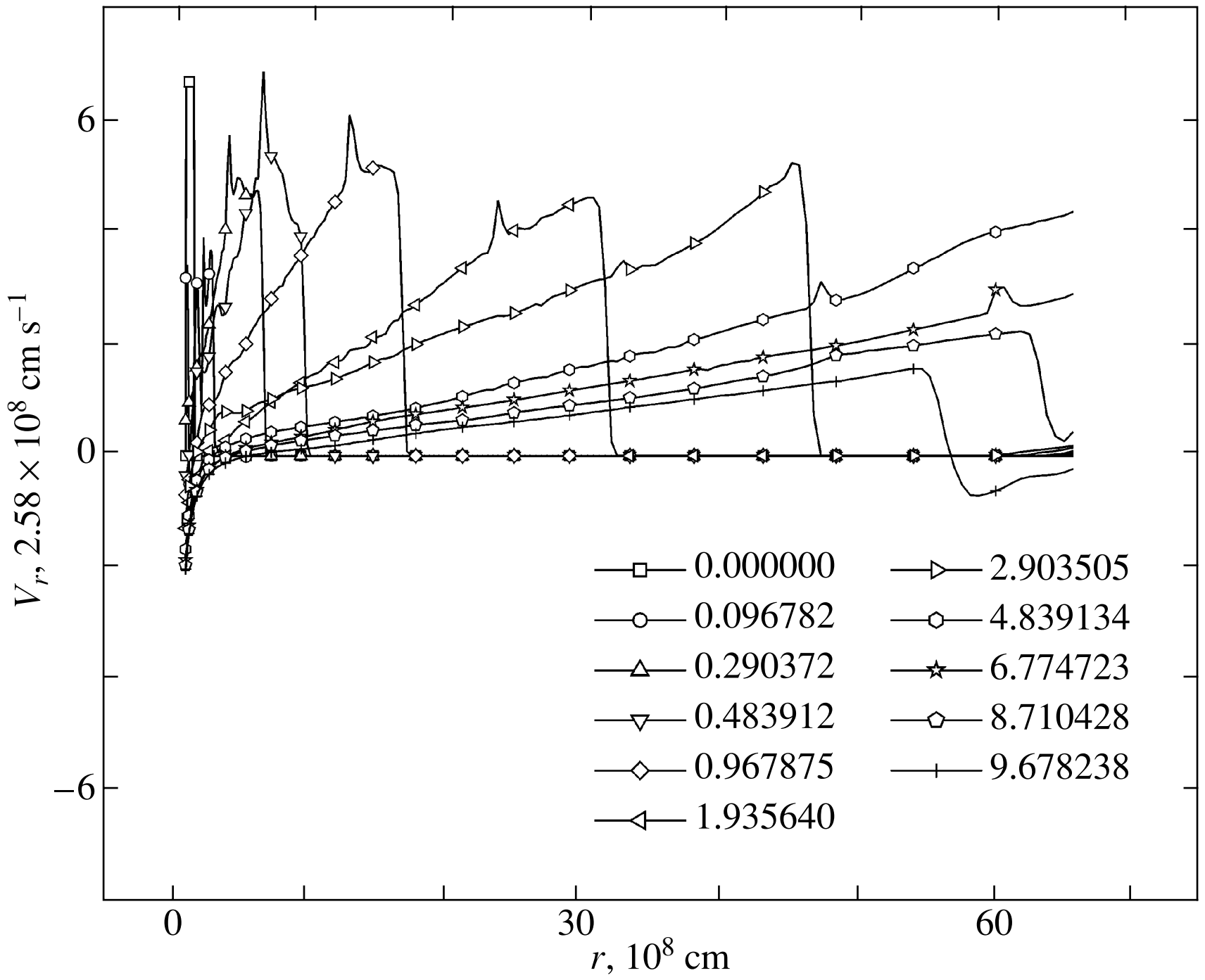

**Fig. 7.** Same as Fig. 1 for data set [7].

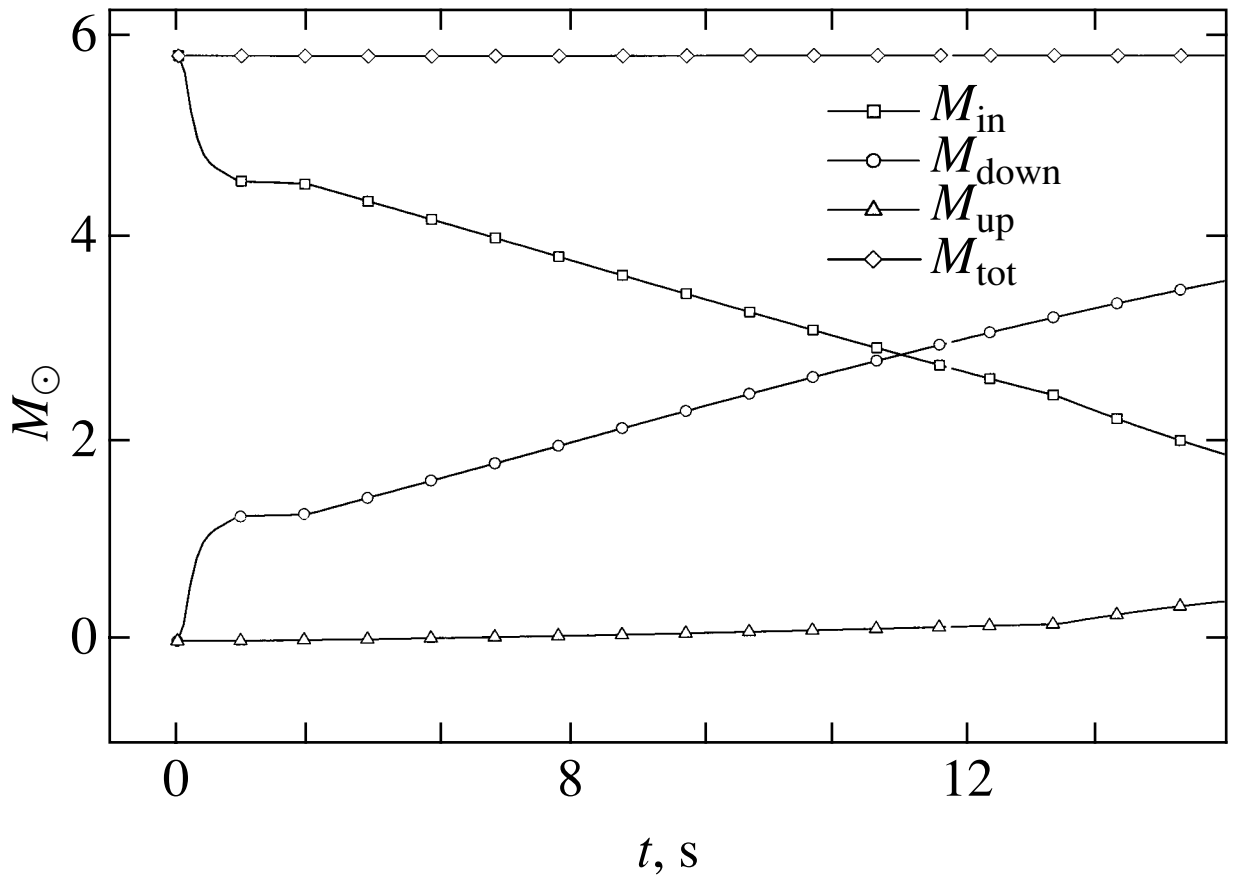

**Fig. 8.** Total mass fluxes through the outer and inner boundaries of the computed region versus time for the calculations with data set [4]; ($M_{in}$ is the total mass of the matter inside the computed region, $M_{down}$ is the total mass flux through the inner boundary, $M_{up}$ is the total mass flux through the outer boundary, and $M_{tot} = M_{in} + M_{down} + M_{up}$ is the total mass of the matter.

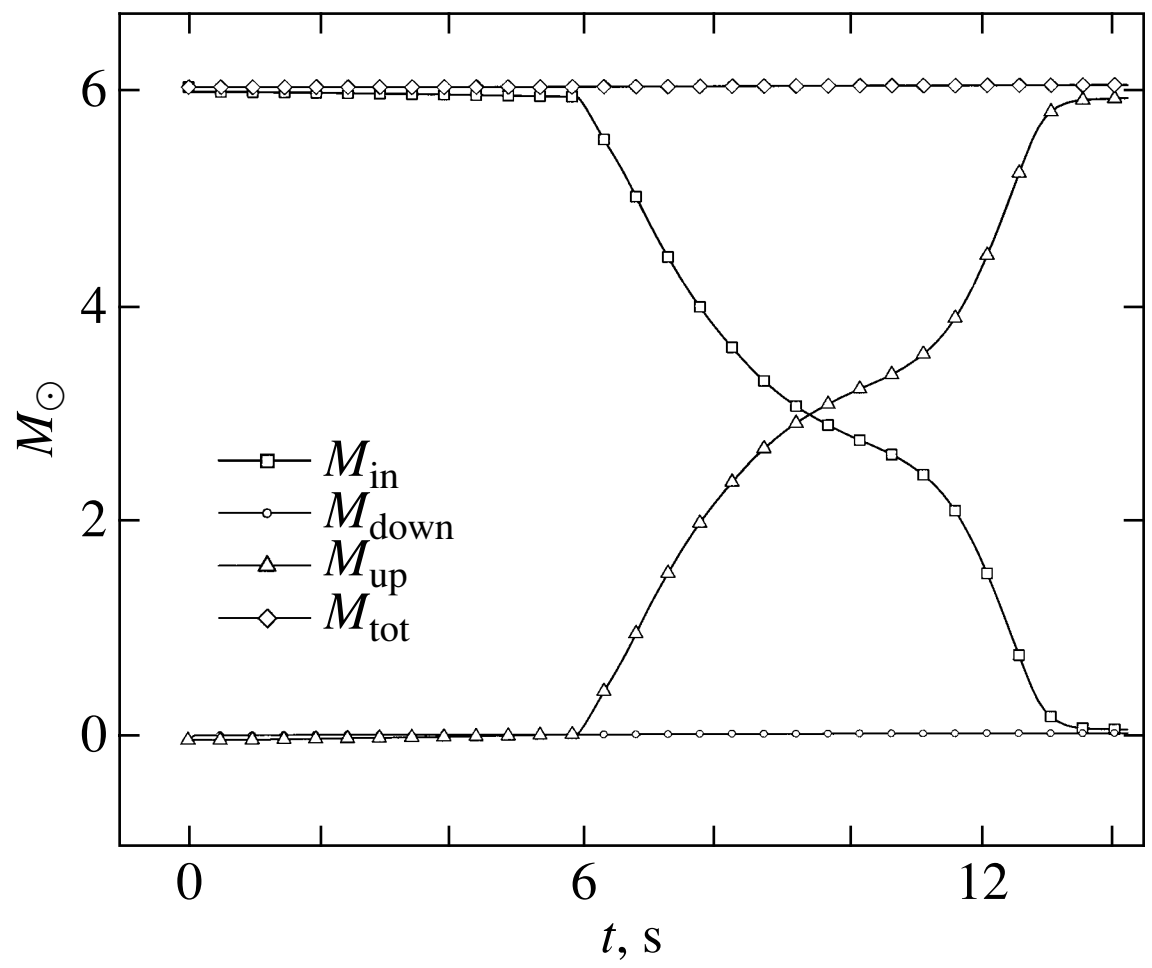

**Fig. 9.** Same as Fig. 8 for data set [6].

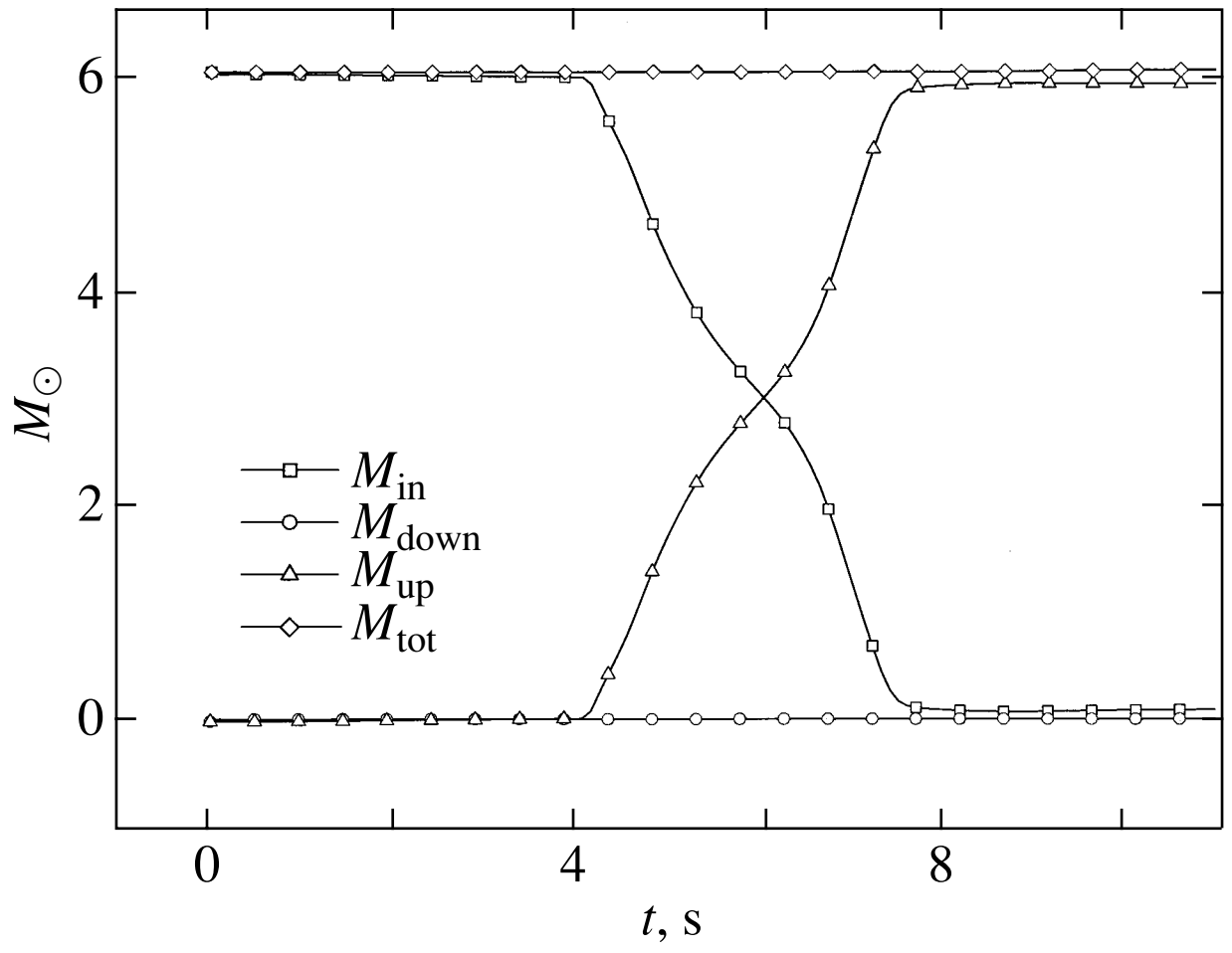

**Fig. 10.** Same as Fig. 8 for data set [7].

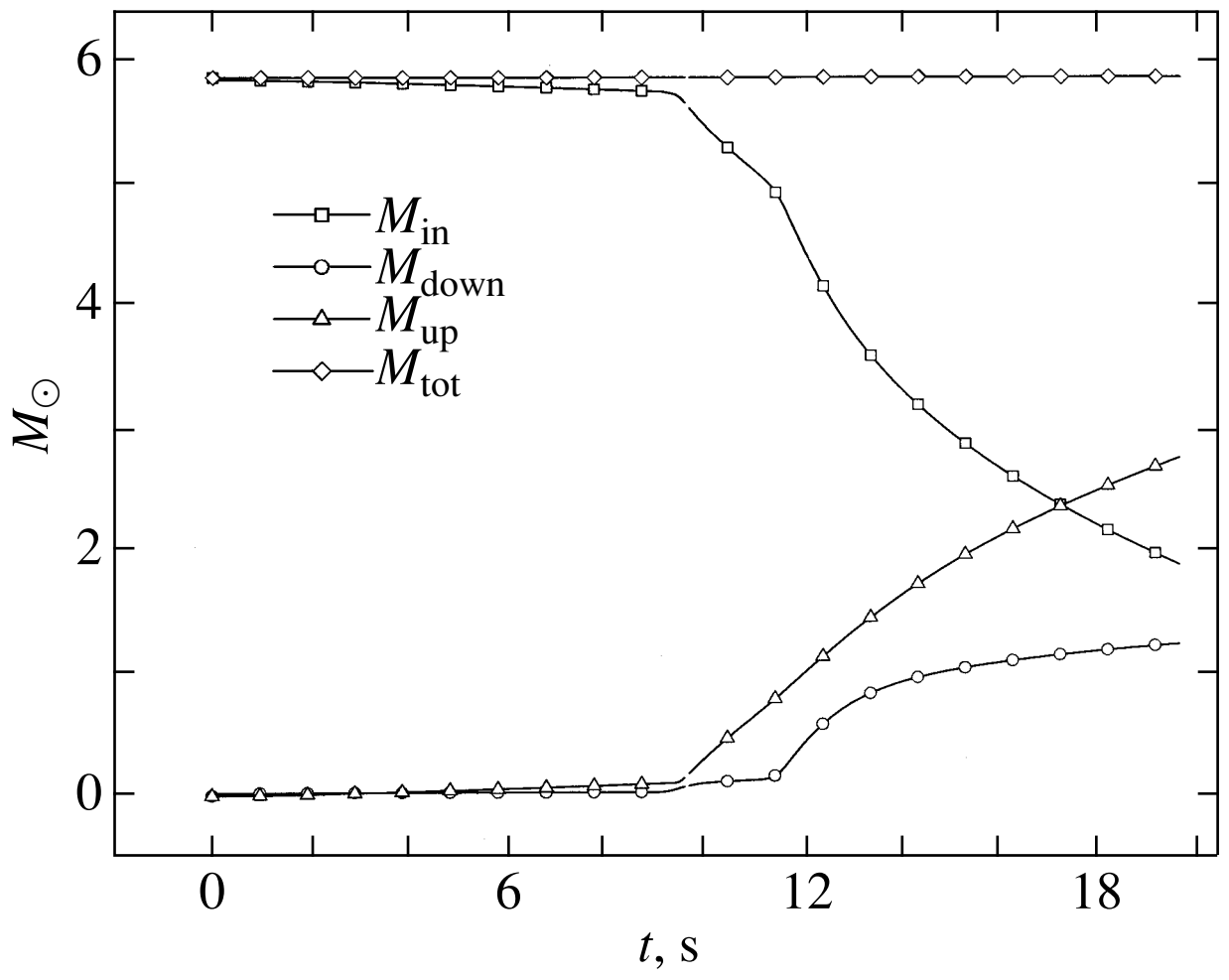

**Fig. 11.** Same as Fig. 8 for data set [5].

In cases [4]–[7], nucleons are recombined into iron nuclei rapidly and in approximately the same time for all cases of initial data, which does not exceed $\sim$1 s. For cases [4]–[7], we also calculated the kinetic energy flux through the outer boundary in radius $r_{\rm max}$. The following values were obtained for the total flux, which may prove to be of use in assessing the possibility of emersion of the outer stellar layers: [4] the total kinetic energy flux through the outer boundary by the completion of the calculation $E_{\rm kin\,out} = 1.80 \times 10^{49}$ erg and the completion time of the calculation $t_{\rm end} = 16.84$ s; [5]: $E_{\rm kin\,out} = 3.87 \times 10^{50}$ erg and $t_{\rm end} = 19.65$ s; [6]: $E_{\rm kin\,out} = 2.23 \times 10^{51}$ erg and $t_{\rm end} = 14.23$ s; [7]: $E_{\rm kin\,out} = 5.46 \times 10^{51}$ erg and $t_{\rm end} = 11.03$ s.

## CONCLUSIONS

Of course, the conclusions regarding the ejection of the entire outer envelope of a massive star during the collapse of its iron core should have been reinforced by hydrodynamic calculations of the propagation of a strong shock wave through the helium shell and through the outer hydrogen–helium shell, certainly if they were preserved during the stellar evolution of this star. However, the velocity of the post-shock matter derived in the calculations with supercritical nucleon-bubble masses, which exceeds $10^9$ cm s$^{-1}$ for a radius of $\sim 6.6 \times 10^9$ cm, is, in turn, appreciably higher than the characteristic free-fall velocity with an inner mass of $6.8M_{\odot}$ ($M =$ $M_{\rm in} + M_{\rm center} - \Delta M_G$), by more than a factor of 2 in case [6] $\left(v_{ff} = \left(\frac{2GM}{r_{\max}}\right)^{1/2} = 5.23 \times 10^8 \text{ cm/s}\right)$. The total outward kinetic energy flux through the outer boundary (see Sect. "Numerical Results") and the estimated recombination energy of the nucleon bubble simulated in the initial conditions of our problem suggest that complete ejection is possible. For the nucleon-bubble mass, we obtain $M_{\rm NB} = 0.996 \times 10^{-2} M_\odot$, so $E_{\rm NB} = M_{\rm NB} \times \frac{8.79 \text{ MeV}}{m_0} = 1.68 \times 10^{50}$ erg. This energy is close in absolute value to the gravitational energy of the outer stellar layers, $\sim 2.5 \times 10^{50}$ erg.

It should be particularly emphasized that our calculations yielded the critical nucleon-bubble mass $(M_{\rm NB})_{\rm crit} = M_{\rm NB}$ ([5]) $= 0.7 \times 10^{-2} M_\odot$ (see case [5]). It should be remembered that here, we deal with the most massive star with $M_{\rm MS} = 25 M_\odot$. For less massive stars, $(M_{\rm NB})_{\rm crit}$ is most likely much lower, although this suggestion needs to be verified by further calculations. Can such nucleon bubbles, which assist in the envelope ejection more than the other gravitational effect of energy losses in the form of a neutrino signal taken into account in our calculations, be formed? In the one-dimensional hydrodynamic theory of gravitational collapse (in many studies with an allowance made for neutrino radiation, including that at the neutrinosphere formation stage, which virtually coincides with the second collapse stage mentioned in the Introduction), a rather thick spherical shell has long been shown to appear above the neutrinosphere, which is semitransparent to neutrino radiation [see Imshennik and Nadyozhin (1982) for a review]. Burrows and Goshy (1993) showed that this shell has a quasi-steady-state structure in the typical case of a frozen accretion wave whose front represents the outer shell boundary. It is important that this shell is composed of a Boltzmann gas of free nucleons and electron–positron plasma. Imshennik (2002) refined such quasi-steady-state solutions and showed its strong convective instability virtually in the entire solution region that encompasses all the possible conditions for gravitational collapse or, to be more precise, its second stage (post-shock accretion). The typical masses in these shells called neutrino PNS coronas are $10^{-3} M_\odot < M_{\rm CPNS} < 10^{-2} M_\odot$; i.e., they are roughly equal to the critical mass $M_{\rm NB}$ derived above. The possibility of such nucleon bubbles being ejected due to the nonlinear development of convection is supported by the well-known considerations given by Burrows *et al.* (1994) based on their calculations of a two-dimensional model for gravitational collapse and by the interesting self-similar solutions of a neutrino wind that drags nucleon bunches into the outer stellar envelopes (Thompson *et al.* 2001). Here, we demonstrated the hydrodynamic effect of the ejection of the outer stellar envelope in the rough one-dimensional approximation. However, to reach the final conclusion regarding the ejection of the entire outer envelope, strictly speaking, requires further calculations of the passage of the shock wave obtained in our calculations through the helium shell and the outer hydrogen–helium shell. Still, it is qualitatively clear that a significant fraction of the kinetic energy accumulated behind the shock front will be spent on overcoming the gravitational binding energy of these outermost layers. For this reason, the resulting mean ejection velocities of the entire envelope can be much lower than the calculated velocities; i.e., the envelope will be essentially emersed rather than ejected.

## ACKNOWLEDGMENTS

We wish to thank A.V. Zabrodin and his coworkers for valuable advice and help in developing the finite-difference method for solving the problem. We are also grateful to S.K. Godunov for attention given to this work. This study was supported by the Russian Foundation for Basic Research (project no. 00-15-96572) and the CRDF (grant no. MO-011-0).

*Translated by V. Astakhov*